\documentclass[a4paper,11pt]{article}
\usepackage{pos}

\newcommand{\FantoPDF}{{\sf FantoPDF} }

\title{Fant\^omas4QCD: pion PDFs with epistemic uncertainties}
 \ShortTitle{Fant\^omas4QCD}

\author*[a]{Aurore Courtoy}
\author[b]{Lucas Kotz}
\author[b]{Pavel Nadolsky}
\author[b]{Fred Olness}
\author[b]{Maximiliano Ponce-Ch\'avez}

\affiliation[a]{Instituto de F\'isica,
  Universidad Nacional Aut\'onoma de M\'exico, \\Apartado Postal 20-364,
  01000 Ciudad de M\'exico, Mexico}

\affiliation[b]{Department of Physics, Southern Methodist University,\\
    Dallas, TX 75275-0175, USA}

\emailAdd{aurore@fisica.unam.mx}

\abstract{In these proceedings, we reiterate and extend the discussion of  the Next-to-Leading Order (NLO) QCD analysis of the charged pion Parton Distribution Function (PDF) obtained within the Fant\^omas4QCD framework. The goal of the Fant\^omas analysis is to quantify the dependence on the parametrization form in global QCD analyses, with a first application to the pion PDFs. We highlight the anti-correlation between the experimentally allowed gluon and sea distributions in the pion, as made apparent by the sampling over parametrization forms. This result, for the sea and gluon sector, illustrates the importance of accounting for epistemic uncertainties in data-driven QCD analyses. In that regard, we further discuss the meaning of the sampling uncertainty and why it is key in phenomenological studies of hadron structure.}

\FullConference{31st International Workshop on Deep Inelastic Scattering (DIS2024)\\
 8--12 April 2024\\
Grenoble, France\\}

\begin{document}
\maketitle

\section{Introduction}

In recent years, QCD analyses of Parton Distribution Functions (PDFs) of the charged pion have been performed using updated data, theoretical input, and statistical methodologies. Both the JAM~\cite{Barry:2021osv} and xFitter~\cite{Novikov:2020snp} collaborations accounted for uncertainties in the pion PDF, employing iterative Monte Carlo and Hessian approaches, respectively. 
This is in contrast with the first analyses, proposed in the 1990s, which focused on the behavior of the central PDF values.  State-of-the-art data then
involved primarily Drell-Yan-like processes, which are mainly sensitive to  quarks and antiquarks, and the interplay between the sea and gluon distributions was highly undetermined. As a result, hypotheses on the sea and gluon distributions shaped the resulting PDFs. The relative contributions of the sea quark and gluon PDFs can be quantified by the net fractions of the pion's momentum they contribute. For instance, SMRS found that a sea momentum fraction as low as $5\%$ at $Q^2=4$ GeV$^2$ poorly fits the NA10 data, with the  $\chi^2$ improving as the sea fraction increased. Conversely, GRV adopted a no-sea and valence-like gluon at the initial scale, an approach that emphasizes the gluon's role.

These diverging results were challenged in recent analyses, with the JAM collaboration asserting that the pion's gluon content was indeed significant, with a momentum fraction $\langle xg\rangle_{\rm JAM}(Q_0)\sim 40\%$.
This result is attributed to the data coverage at small $x$ from leading-neutron deep inelastic scattering, which the JAM collaboration chose to include, despite some model dependence introduced by the pion flux.
\\

The Fant\^omas4QCD project\footnote{The Fant\^omas4QCD project is led by members of the CTEQ-TEA collaboration.} has proposed the first determination of the pion PDFs with {\sf epistemic} uncertainties~\cite{Kotz:2023pbu}. Beyond the propagation of statistical uncertainties, these account for sampling accuracy, with sampling done over PDF parametrization forms. 
By introducing the concept of a {\sf metamorph}, we generate varying and flexible parametrizations, in the form of Bézier curves, used as trial functions in the optimization process. We tested 100 settings of the {\sf metamorph} and found about $20\%$ of acceptable solutions to the analysis of Drell-Yan, prompt-photon, and one leading-neutron data set. With these solutions, we concluded that the sea and gluon distributions remain weakly constrained solely by the experimental measurements and anti-correlated,  
contradicting claims of precise momentum fractions for the gluon PDF in the pion.\\

In what follows, we will summarize our methodology and the resulting PDFs with their respective momentum fractions.

\begin{figure}[t]
\begin{center}
    \includegraphics[width=0.475\columnwidth]{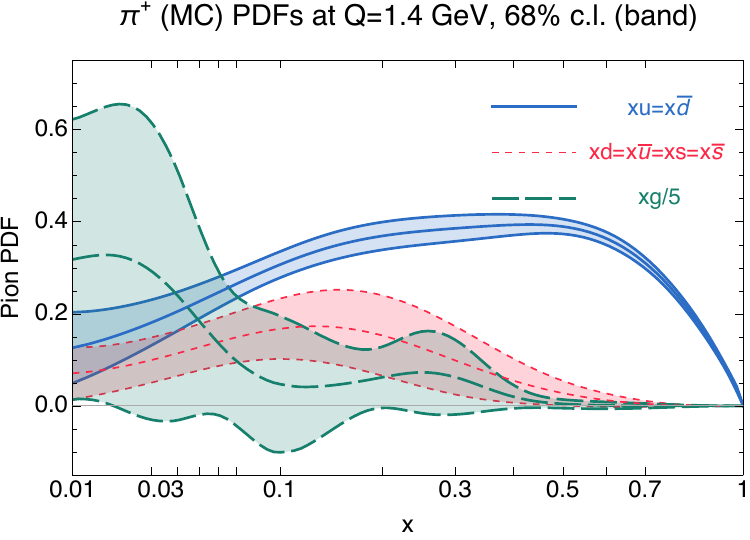}
    \includegraphics[width=0.475\columnwidth]{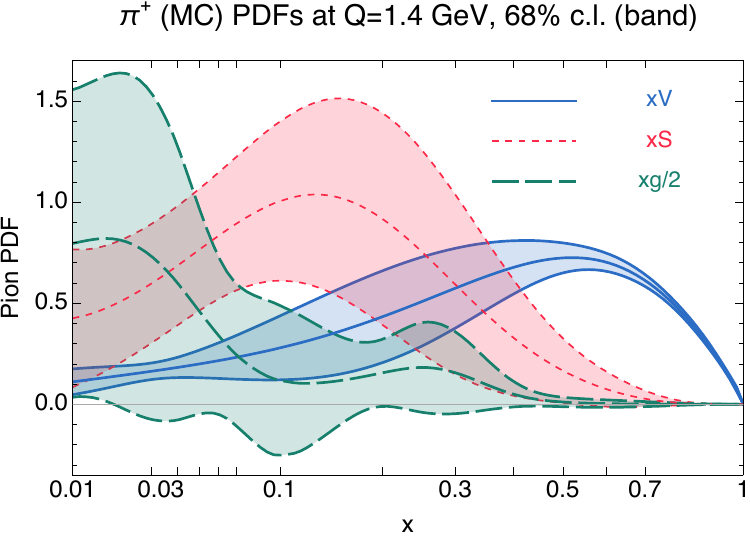}
    \caption{The Fant\^oPDFs for the pion, at $Q_0=\sqrt{1.9}$ GeV. Left: the physical flavor basis, right: valence, sea and gluon distributions. Notice that Fig.~9 of Ref.~\cite{Kotz:2023pbu} is mislabeled.}
    \label{fig:fantofl}
\end{center}
\end{figure}

\section{Fant\^omas4QCD}

The purpose of our program is to explore the sampling uncertainty in PDF analyses. 
To find PDFs as functions of the variable $x$, the fraction of the parent’s longitudinal momentum carried by the probed parton, they are fitted to experimental measurements, perhaps supplemented by theoretical inputs from lattice QCD or other techniques.
In most practical cases, inverse problems associated with such fits  do not have a unique solution; {\it they are ill-posed}. Multiple functions can fulfill all requirements for goodness-of-fit criteria satisfactorily. In other words, the current state of the data, with its corresponding aleatoric uncertainty, does not allow for disentangling between the various solutions. Hence, to account for the full space of solutions representatively, we sample over a  
collection of functions of $x$ that describe the data equally well.

This methodological step follows from a previous analysis by some of the authors of these proceedings, which argued for the importance of replicable PDF sampling~\cite{Courtoy:2022ocu}. Exploiting representative sampling further, we have designed {\it metamorphs}  -- PDF parametrizations  
composed of two factors: a carrier function that ensures the asymptotics at the endpoints $x \to $ 0 and 1, and a modulator function that provides flexibility over the entire $x$ range,
\begin{eqnarray}
    x\,f_i(x, Q_0^2)&=& F_i^{car} \times F_i^{mod}\nonumber\\
    &=&A_i x^{B_i} (1-x)^{C_i} \times \left[ 1+ {\cal B}^{(N_m)}(y(x)) \right].
    \label{eq:xfpion} 
\end{eqnarray}
The difference between a classical polynomial fit and a metamorph as described in Eq.~(\ref{eq:xfpion}) is that the Bézier curve ${\cal B}^{(N_m)}(y)$ is uniquely determined by its values $P_j$ at control points $x_j$, {\it i.e.}, 
$P_j\equiv {\cal B}^{(N_m)}(y(x_j))$, with $y(x)$ a function of $x$. Using $N_m+1$ such control points, the vector of coefficients of the polynomial can uniquely be determined by a simple matrix equation (see Ref.~\cite{Kotz:2023pbu} and references therein). The sub-index $i$ refers to $V, S$ and $ g$, with $V, S$ defined by
\begin{eqnarray}
u^{\pi^+}=\bar d^{\pi^+}=\bar{u}^{\pi^-}=d^{\pi^-} = V/2  + S/6, 
&& \bar{u}^{\pi^+}=\bar{d}^{\pi^-}=s^{\pi^{\pm}}={\bar s}^{\pi^{\pm}} = S/6.
\label{S}
\end{eqnarray}

By varying the settings of the {\sf metamorph} $\{ P_j, y(x), N_m\}$, many parametrizations can be generated on the fly. Filtering the acceptable solutions through soft priors, handled manually, we are left with a set of solutions from which the 5 most diverse shapes at $Q_0$ are kept. These 5 solutions, each provided as a Hessian error set with $\Delta \chi^2=1$, are combined using the METAPDF technique~\cite{Gao:2013bia}. This involves generating $N_{\rm repl}$ Monte Carlo replicas per set and combining all 5 sets in an unweighted manner.
Hence, the final set of the Fant\^oPDFs for the pion, depicted in Fig.~\ref{fig:fantofl}, represents a solution to a Bayesian inverse problem that now accounts for both {\sf aleatoric} and {\sf epistemic} uncertainties. Fant\^omas4QCD provides an alternative to the neural network  methodology for solving Bayesian inverse problems (see, {\it e.g.}, Ref.~\cite{DelDebbio:2022tsj}).

\section{Anticorrelation of gluon and sea PDFs}

\begin{figure}[t]
\begin{center}
    \includegraphics[width=0.65\columnwidth]{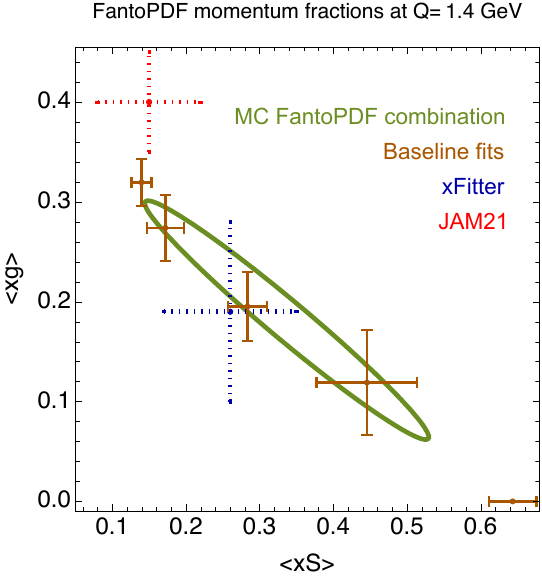}
    \caption{Momentum fractions for the sea and gluon PDFs obtained in the five baseline fits (orange-brown) and the MC \FantoPDF combination~\cite{Kotz:2023pbu}   (green ellipse) at $Q_0$. JAM21 \cite{Barry:2021osv} and {\sf xFitter} \cite{Novikov:2020snp} results are displayed in red and blue, respectively. }
    \label{fig:moment_frac_corr}
\end{center}
\end{figure}

 The Fant\^omas pion PDFs were determined
 from the analysis of Drell-Yan (DY) pair production by $\pi^-$ scattering on a tungsten target by
 E615 (140 data points) and NA10 (140 data points), 
 prompt-photon ($\gamma$) production in $\pi^-$ and  $\pi^+$ scattering on a tungsten target by WA70 (99 data points) -- the baseline of the xFitter analyses-- with the addition of $29$~leading-neutron data points from the H1 collaboration. 
The prompt photon as well as the leading-neutron data are more sensitive to gluon and sea quark distributions. While the area under the valence PDF is constrained by the number density sum rule, the sea and gluon distribution are only constrained in a sum that involves all partons, through the momentum sum rule
\begin{eqnarray}
    \sum_{q=V, S, g}\langle x q(Q^2)\rangle&=&1\, ,\\
    {\rm with} &&\langle x q(Q^2)\rangle=\int_0^1 dx\, x\, f_{1, \pi}^{q}(x, Q^2)\, . \nonumber
\end{eqnarray}
The valence contribution to the momentum sum rule was found to hardly change with the different parametrizations, implying an anti-correlation between the momentum fractions carried by the sea and the gluon PDFs.  
At the initial scale $Q^2_0 =1.9 \mbox{ GeV}^2$ of DGLAP evolution, we have found solutions with essentially zero gluon PDF and high sea PDF and, conversely, rapidly growing gluon PDF at small $x$ and low sea PDF.
We confirm that the NA10 data prefer a low gluon at NLO, while no stable sensitivity pattern appears for other data sets. 
These results are reflected in the size of the uncertainties in the Fant\^oPDF combination  in Fig.~\ref{fig:fantofl} as well as in the distribution of the momentum fractions between the sea and the gluon in Fig.~\ref{fig:moment_frac_corr}. 
In the latter figure, we show the momentum fractions for the 5 baseline fits as well as the correlation ellipse that results from the FantôPDF combination. These results are compared to the JAM and xFitter predictions. Despite sampling over parametrizations, we did not find gluon momentum fractions as high as $40\%$. The differences between our analysis and JAM's may lie in the choice of data, the treatment of the  model dependence for the leading-neutron DIS data, as well as methodological choices. These issues are avenues for future investigation.

\section{Conclusions}

In these proceedings, we summarized the  Fant\^omas4QCD project ---a phenomenological NLO analysis 
of charged pion PDFs implemented in the xFitter program that investigates the PDF parametrization dependence. The sampling over the space of PDF solutions is an important factor in the total uncertainty on PDFs that was not considered by previous pion studies. The magnitude of uncertainty tolerated by experimental measurements affects our knowledge on meson structure -- how much glue do we expect in a pion at a given scale?
For now, the answer is that there is a large uncertainty on the momentum fraction carried by the gluons at DIS-like scales, for which factorization theorems apply (see Ref.~\cite{Courtoy:2020fex}). 

Further study of the parametrization sampling will be pursued for the determination of the proton PDF in the context of the next-generation CTEQ-TEA PDFs. The collaboration is engaged in exploring epistemic PDF uncertainties in the near future.

\section*{Acknowledgments}

This study has been financially supported by 
the Inter-American Network of Networks of QCD Challenges,
a National Science Foundation AccelNet project, 
by CONACyT--Ciencia de Frontera 2019 No.~51244 (FORDECYT-PRONACES),
by the U.S.\ Department of Energy under Grant No.~DE-SC0010129,
and 
by the  U.S.\ Department of Energy, Office of Science, Office of Nuclear Physics, 
within the framework of the Saturated Glue (SURGE) Topical Theory Collaboration.
AC and MPC were further supported by the UNAM Grant No. DGAPA-PAPIIT IN111222.


\begin{thebibliography}{99}


\bibitem{Barry:2021osv}
P.~C.~Barry \textit{et al.} [Jefferson Lab Angular Momentum (JAM)],
Phys. Rev. Lett. \textbf{127} (2021) no.23, 232001
doi:10.1103/PhysRevLett.127.232001
[arXiv:2108.05822 [hep-ph]].

\bibitem{Novikov:2020snp}
I.~Novikov, H.~Abdolmaleki, D.~Britzger, A.~Cooper-Sarkar, F.~Giuli, A.~Glazov, A.~Kusina, A.~Luszczak, F.~Olness and P.~Starovoitov, \textit{et al.}
Phys. Rev. D \textbf{102} (2020) no.1, 014040
doi:10.1103/PhysRevD.102.014040
[arXiv:2002.02902 [hep-ph]].

\bibitem{Kotz:2023pbu}
L.~Kotz, A.~Courtoy, P.~Nadolsky, F.~Olness and M.~Ponce-Chavez,
Phys. Rev. D \textbf{109} (2024) no.7, 074027
doi:10.1103/PhysRevD.109.074027
[arXiv:2311.08447 [hep-ph]].


\bibitem{Courtoy:2022ocu}
A.~Courtoy, J.~Huston, P.~Nadolsky, K.~Xie, M.~Yan and C.~P.~Yuan,
Phys. Rev. D \textbf{107} (2023) no.3, 034008
doi:10.1103/PhysRevD.107.034008
[arXiv:2205.10444 [hep-ph]].

\bibitem{Gao:2013bia}
J.~Gao and P.~Nadolsky,
JHEP \textbf{07} (2014), 035
doi:10.1007/JHEP07(2014)035
[arXiv:1401.0013 [hep-ph]].

\bibitem{DelDebbio:2022tsj}
L.~Del Debbio,
SciPost Phys. Proc. \textbf{15} (2024), 028
doi:10.21468/SciPostPhysProc.15.028
[arXiv:2211.00977 [hep-lat]].

\bibitem{Courtoy:2020fex}
A.~Courtoy and P.~M.~Nadolsky,
Phys. Rev. D \textbf{103} (2021) no.5, 054029
doi:10.1103/PhysRevD.103.054029
[arXiv:2011.10078 [hep-ph]].




\end{thebibliography}
\end{document}